\documentclass[10pt, twocolumn, a4page]{ileapr}
\usepackage{graphicx}
\usepackage{times}    
\usepackage{amsmath}  
 \newcommand{\mb}[1]{\mathbf{#1}}
 
 \newcommand{\mr}[1]{\mathrm{#1}}

 \newcommand{\D}{\Delta}
\input{ileapr.mac}
\begin{document}

\title{
 \vskip -15mm
         {\large Annual Progress Report 2003, Institute of Laser Engineering, Osaka University (2004), pp.151-154}\\
 \vskip -5mm
\rule{170mm}{0.2mm}
 \vskip  2mm
\bf A New Time-Reversible Integrator for Molecular Dynamics Applications}
\author{Vasilii Zhakhovskii}
\maketitle
\thispagestyle{empty}

\section*{ INRODUCTION }

Nowadays Molecular Dynamics (MD) approach is widely applied in many areas of physics, chemistry and biochemistry. The
MD is based on solution of second order differential equations of motion, this is why the integration algorithm is a
cornerstone of the MD method \cite{Allen}. The Newtonian equations of motion are time-reversible and it would be
reasonable to preserve this essential property in our integration schemes. Since 1990 there are many nice symplectic
integrators were invented \cite{Omelyan}, mainly in force-gradient form but none with higher-order gradient operators.
In this short report we derive a new time-reversible explicit integrator on the basis of second order Tailor expansion
of force. There is good reason to think the new method will be easy-to-use for MD and, possibly, celestial mechanics
applications.

\section*{ IDEA OF A NEW METHOD }

Consider the second order differential equation:
\begin{equation}
                \ddot{x} = f(x)                      \label{eq:idea1}
\end{equation}
It is useful to introduce the function $s(t)$:
\begin{eqnarray}
   s(t) =
\left\{ \begin{array}{cr}
        1+t/h, &\quad -h<t<0 \\
        1-t/h, &\quad  0<t<h
   \end{array} \right.                               \label{eq:idea2}
\end{eqnarray}
By using Eq.(\ref{eq:idea1}) one may integrate by parts the integral
\begin{equation*}
\begin{split}
\int_{-h}^h f(x) s(t) dt & = \int_{-h}^0 f(x) s(t) dt+\int_{0}^h f(x) s(t) dt= \\
                         & = (\dot{x} s - x \dot{s})|_{-h}^0 + (\dot{x} s - x \dot{s})|_{0}^h ,
\end{split}
\end{equation*}
and finally
\begin{equation}
x(-h)-2x(0)+x(h) = h \int_{-h}^h f(x) s(t) dt .      \label{eq:idea3}
\end{equation}
A proper approximation of the function $f(x)$ in the Eq.(\ref{eq:idea3}) within the segment $t\in [-h,h]$ may
give us difference schemes for numerical integration of Eq.(\ref{eq:idea1}). For instance, by assuming
$f(x)\approx f(x(0))$, the exact formula (\ref{eq:idea3}) immediately gives the explicit Verlet integrator:
\begin{equation}
x(-h)-2x(0)+x(h) = f(x(0)) h^2   \label{eq:idea4}
\end{equation}
Because Eq.(\ref{eq:idea4}) is symmetric for $t\pm h$ the Verlet method is time-reversible.
In the case of time quadratic approximation of $f(x(t))$ the implicit Stoermer method is derived
from Eq.(\ref{eq:idea3}):
\begin{equation}
\begin{split}
 & x(-h)-2x(0)+x(h) = \\
 & = (f(x(-h))+10f(x(0))+f(x(h)) ) h^2/12     \label{eq:idea5}
\end{split}
\end{equation}

It is convenient to introduce notations:
\begin{eqnarray}
       x_0 & = & x(0), \quad  x_h = x(h), \quad  x_{-h} = x(-h)                  \nonumber        \\
\delta x_0 & = & x_0 - x_{-h}, \quad  \delta x_h = x_h - x_0                     \nonumber         \\
      v(t) & = & \dot{x}(t), \quad  a(t) = \ddot{x}(t), \quad  b(t) = \dot{a}(t) \label{eq:idea5a}  \\
       f_0 & = & f(x_0), \quad  f_h = f(x_h), \quad  f_{-h} = f(x_{-h})          \nonumber           \\
     f'(x) & = & df(x)/dx, \quad  f''(x)= d^2f(x)/dx^2                           \nonumber
\end{eqnarray}
Thus the Verlet method (\ref{eq:idea4}) is given by
\begin{equation}
\delta x_h = \delta x_0 + a_0h^2 + O(h^4)                                  \label{eq:idea4a}
\end{equation}

Expand $f(x)$ in the vicinity of $x=x_0$  and $x(t)$ at a point $t=0$ in a Taylor series
\begin{equation}
f(x) = f_0 + f'_0 (x-x_0) + f''_0(x-x_0)^2/2 + f'''_0 (x-x_0)^3/6 + ..     \label{eq:idea6}
\end{equation}
\begin{equation}
x-x_0 = v_0 t + a_0 t^2/2 + b_0 t^3/6 + ..                                 \label{eq:idea7}
\end{equation}

Substitute $(x-x_0)$ from Eq.(\ref{eq:idea7}) to Eq.(\ref{eq:idea6}) and hold only the even terms so
one may find that the function $f_{even}$ along the trajectory of motion around $t=0$ is
\begin{equation}
f_{even}(t) = f_0 + (a_0f'_0 + v_0^2 f''_0) t^2/2 + O(t^4)                   \label{eq:idea8}
\end{equation}
Substitute Eq.(\ref{eq:idea8}) to Eq.(\ref{eq:idea3}) and integrate the latter:
$$
h\int_{-h}^h t^2 s(t) dt = h^4/6
$$
and
$$
     c_0 \equiv a_0f'_0 + v_0^2 f''_0
$$
\begin{equation}
\delta x_h = \delta x_0 + a_0h^2 + c_0 h^4/12 + O(h^6)             \label{eq:idea9}
\end{equation}
The explicit integrator (\ref{eq:idea9}) is time-reversible likewise the Verlet method. The important difference of
Eq.(\ref{eq:idea9}) from Verlet formula (\ref{eq:idea4a}) is a velocity and acceleration dependencies in the $c_0$
coefficient.

 \underline{How to evaluate velocity $v_h$ ?}

{\it 1st way}.
By using the same approach as above for derivation of (\ref{eq:idea9}) one may deduce a formula
\begin{equation}
v_h = v_{-h} + \int_{-h}^h f(x) dt .                               \label{eq:idea10}
\end{equation}
Therefore the time-reversible velocity formula is
\begin{equation}
 v_h = v_{-h} + f_0 2h + (a_0f'_0 + v_0^2 f''_0) h^3/3 + O(h^5)    \label{eq:idea11}
\end{equation}
The main disadvantage of Eq.(\ref{eq:idea11}) is a poor accuracy.

{\it 2nd way}.
Assume that in the vicinity of $t=0$ the $x(t)$ can be represented by polynomial interpolation:
$$
x(t)= x_0+v_0t+a_0 t^2/2 +c_3t^3+c_4t^4+c_5t^5+c_6t^6
$$
By using the known positions, velocities, and accelerations at points $t=[-h,0,h]$ (see Table 1) one can derive
a time-reversible formula for evaluation of velocity at time point $t=h$:
\begin{table}
             \caption{ The source data for the time-reversible interpolation formula of velocity (\ref{eq:idea12}) }
 \begin{center}
 \begin{tabular}{|c||c|c|c|}
  \hline
        $ t $ &  $-h $         &  $  0  $     &   $  h $  \\
  \hline
  \hline
        $ x $ &  $x_{-h}$      &  $x_0$       &   $x_h$    \\
  \hline
        $ v $ &  $v_{-h}$      &  $v_0$       &            \\
  \hline
        $ a $ &  $a_{-h}$      &  $a_0$       &   $a_h$    \\
    \hline
 \end{tabular}
 \end{center}
\end{table}
\begin{equation}
 v_h = v_{-h} + (\delta x_h-\delta x_0)\frac{8}{3h} + (a_{-h}+a_{h}-8a_0)\frac{h}{9} + O(h^6)  \label{eq:idea12}
\end{equation}
Eq.(\ref{eq:idea12}) is more accurate than Eq.(\ref{eq:idea11}), but it requires the knowledge of position and
acceleration at $t=h$. The general scheme of the new integrator on the basis of formulae
(\ref{eq:idea9},\ref{eq:idea12}) is presented in the Table 2. The first time-step has to be done by using another
integrator.

{\it 3rd way}. One may obtain even more precise velocity formula by using correction term $c_0$ of Eq.(\ref{eq:idea9}):
\begin{equation}
\begin{split}
   v_h = v_{-h} & + (\delta x_h-\delta x_0)\frac{48}{13h} + (a_{-h}+a_{h}-24a_0)\frac{h}{13}- \\
                & - c_0\frac{8h^3}{12\cdot 13} + O(h^8)                                       \label{eq:idea12a}
\end{split}
\end{equation}
It should be noted the precision of estimate of velocity by Eq.(\ref{eq:idea12a}) exceeds the precision of coordinate
evaluation, therefore the energy conservation can not be improved considerably by this way.


\begin{table}
             \caption{ Algorithm of integrator based on formulae (\ref{eq:idea9},\ref{eq:idea12a}). Here
             $K$ is a time-step number and $L$ is a logical step number within the given time-step. The first
             time-step $0_0 \to 1_0$ has to be done by using another integrator.  }
 \begin{center}
 \begin{tabular}{|c||c|c|c|c|c|c|c|}
  \hline
  $ K_L $       &  $0_0$  &  $1_0 $  & $2_1$ & $2_2$ & $2_3$ & $2_4$  & $2_0$  \\
  \hline
  \hline
        $ x $   &  $x_0$  &  $x_1$   & $x_2$ &       &       &        & $x_2$     \\
  \hline
        $ v $   &  $v_0$  &  $v_1$   &       &       & $v_2$ &        & $v_2$     \\
  \hline
        $ a $   &  $a_0$  &  $a_1$   &       & $a_2$ &       &        & $a_2$     \\
  \hline
        $ c $   &         &  $c_1$   &       &       &       & $c_2$  & $c_2$    \\
  \hline
 \end{tabular}
 \end{center}
\end{table}

\section*{ VECTOR FORMULAE }

Again let us consider the second order differential equation in vector notation:
\begin{equation}
   \ddot{\mb{r}} = \mb{f(r)}/m = \mb{a}                                              \label{eq:3D1}
\end{equation}
Eqs.(\ref{eq:idea2},\ref{eq:idea6},\ref{eq:idea7}) can be rewritten as vector equations:
\begin{equation}
   \delta \mb{r}_h = \delta \mb{r}_0 +\frac{h}{m}\int_{-h}^h \mb{f(r)} s(t) dt .     \label{eq:3D2}
\end{equation}

\begin{equation}
   \mb{f(r}_0+\D\mb{r)} = \mb{f}_0 +(\D\mb{r}\cdot \nabla)\;\mb{f}_0 +
                                    (\D\mb{r}\cdot \nabla)^2\; \mb{f}_0/2 + ...        \label{eq:3D3}
\end{equation}

\begin{equation}
   \D\mb{r} = \mb{v}_0 t + \mb{a}_0 t^2/2 + \mb{b}_0 t^3/6 + ...                     \label{eq:3D4}
\end{equation}
As it was for Eq.(\ref{eq:idea8}), substitute $\D\mb{r}$ from Eq.(\ref{eq:3D4}) to Eq.(\ref{eq:3D3}) and
hold only the even terms
\begin{equation}
\begin{split}
   \mb{f(r}_0+\D\mb{r)}_{even} = \mb{f}_0  & + \left[(\mb{a}_0 \cdot \nabla)\;\mb{f}_0 +
            (\mb{v}_0 \cdot \nabla)^2 \;\mb{f}_0 \right]\frac{t^2}{2} + \\
                                           & + O(t^4)        \label{eq:3D5}
\end{split}
\end{equation}
and finally
\begin{equation}
\begin{split}
  & \mb{c}_0 = \left[(\mb{a} \cdot \nabla)\;\mb{f} + (\mb{v} \cdot \nabla)^2\; \mb{f} \right]_0/m \\
  & \delta \mb{r}_h = \delta \mb{r}_0 + \mb{a}_0 h^2 + \mb{c}_0 h^4/12 + O(h^6)    \label{eq:3D6}
\end{split}
\end{equation}
For evaluation of velocity one may again use Eq.(\ref{eq:idea12a}) and the general scheme of integration (see Table 2).

\underline{1st example}\\
Consider the forces depending only on distance $r=|\mb{r}|$
$$
   \mb{f(r)} = f(r)\; \mb{r}
$$
We need the following vector identities
\begin{equation}
   (\mb{w}\cdot \nabla)\mb{f} = f(r)\;\mb{w} + \frac{f'}{r}(\mb{w\cdot r)\; r}         \label{eq:1Ex1}
\end{equation}

\begin{equation}
\begin{split}
   (\mb{w}\cdot \nabla)^2\mb{f} & = 2\frac{f'}{r}(\mb{w\cdot r)\; w} + \frac{f'}{r}(\mb{w\cdot w)\; r} + \\
          & + \left(\frac{f''}{r^2} - \frac{f'}{r^3}\right) (\mb{w\cdot r)}^2\; \mb{r}    \label{eq:1Ex2}
\end{split}
\end{equation}
where $\mb{w}=\mr{const}$.

Applying Eqs.(\ref{eq:1Ex1},\ref{eq:1Ex2}) to the general integration formula Eq.(\ref{eq:3D6})
\begin{equation}
\begin{split}
    \mb{c}_0 & = \left[f(r)\;\mb{a}+ 2\frac{f'}{r}(\mb{v\cdot r)\; v}+ \frac{f'}{r}(\mb{v}^2 + \right. \\
             & + \left. (\mb{a\cdot r))\; r}+
 \left(\frac{f''}{r^2} - \frac{f'}{r^3}\right) (\mb{v\cdot r)}^2\;\mb{r}\right]_0/m     \label{eq:1Ex3}
\end{split}
\end{equation}
\begin{equation}
   \delta \mb{r}_h = \delta \mb{r}_0 + \mb{a}_0 h^2 + \mb{c}_0 h^4/12                    \label{eq:1Ex4}
\end{equation}

\underline{2nd example}\\
Let us consider the Molecular Dynamics system of $N$ particles with pair-wise interaction.
In the vector notation $N$ equations of motion are given by
\begin{equation}
   m_i \ddot{\mb{r}}_i = \sum_{j\ne i} f(r_{ij})\; \mb{r}_{ij}                   \label{eq:2Ex1}
\end{equation}
Represent the relative positions $\mb{r}_{ij}$ in a Taylor series about $t=0$
\begin{equation}
   \D\mb{r}_{ij} = \mb{r}_{ij}(t) -\mb{r}_{ij}(0) = (\mb{v}_i-\mb{v}_j)_0\; t +
                   (\mb{a}_i-\mb{a}_j)_0\; t^2/2 + ..                            \label{eq:2Ex2}
\end{equation}
and expand the pair-wise forces $\mb{f}_{ij}$ in the vicinity of $\mb{r}_{ij}(0)$
\begin{equation}
\begin{split}
   \mb{f}_{ij}(\mb{r}_{ij}(0)+\D\mb{r}_{ij}) & = \mb{f}_{ij,0} +(\D\mb{r}\cdot \nabla)_{ij}\;\mb{f}_{ij,0}+\\
          & + (\D\mb{r}\cdot \nabla)^2_{ij}\; \mb{f}_{ij,0}/2 + ..                         \label{eq:2Ex3}
\end{split}
\end{equation}
By substitution $\D\mb{r}_{ij}$ from Eq.(\ref{eq:2Ex2}) to Eq.(\ref{eq:2Ex3}) one may obtain
\begin{equation}
\begin{split}
  & \delta \mb{r}_{i,h} = \delta \mb{r}_{i,0} + \sum_{j\ne i} \frac{h^2}{m_i}\mb{f}_{ij,0}  +
      \frac{h^4}{12m_i}\left[  \right. \\
  & \left.   ((\mb{a}_i-\mb{a}_j) \cdot \nabla_{ij})_0 \;\mb{f}_{ij,0}  +
   ((\mb{v}_i-\mb{v}_j) \cdot \nabla_{ij})^2_0 \; \mb{f}_{ij,0} \right]            \label{eq:2Ex4}
\end{split}
\end{equation}

Then, by substitution $\mb{a}_{ij}=(\mb{a}_i-\mb{a}_j)$ and so on into Eqs.(\ref{eq:1Ex3},\ref{eq:1Ex4})
it is easy to derive the final formula:
\begin{equation}
\begin{split}
   & \mb{c}_{ij,0} = \left[f(r)\;\mb{a} +  2\frac{f'}{r}(\mb{v\cdot r)\; v} +  \right. \\
   & + \left. \frac{f'}{r}(\mb{v}^2 + (\mb{a\cdot r))\; r}+
              \frac{rf''-f'}{r^3} (\mb{v\cdot r)}^2\;\mb{r}\right]_{ij,0} /m_i        \label{eq:2Ex5}
\end{split}
\end{equation}
\begin{equation}
   \delta \mb{r}_{i,h} = \delta \mb{r}_{i,0} + \mb{a}_{i,0} h^2  +
                         \sum_{j\ne i} \mb{c}_{ij,0}\; h^4/12                          \label{eq:2Ex6}
\end{equation}

\section*{ MOLECULAR DYNAMICS TESTING }

\begin{figure}
 \begin{center}
  \includegraphics [width=\columnwidth] {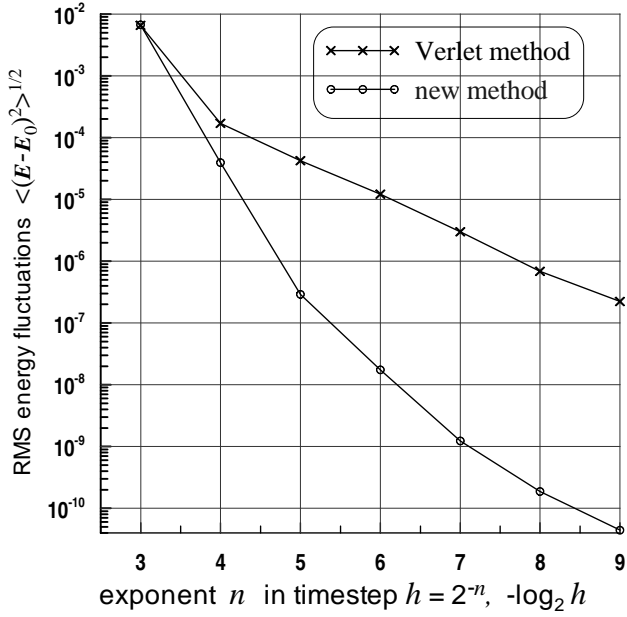}
 \vskip 3mm
 \caption{\label{fig:Err} Time-averaging numerical accuracy of two methods as a function of time-step
                          at equal numbers of MD simulation steps.   }
 \end{center}
\end{figure}

\begin{figure}
 \begin{center}
  \includegraphics [width=\columnwidth] {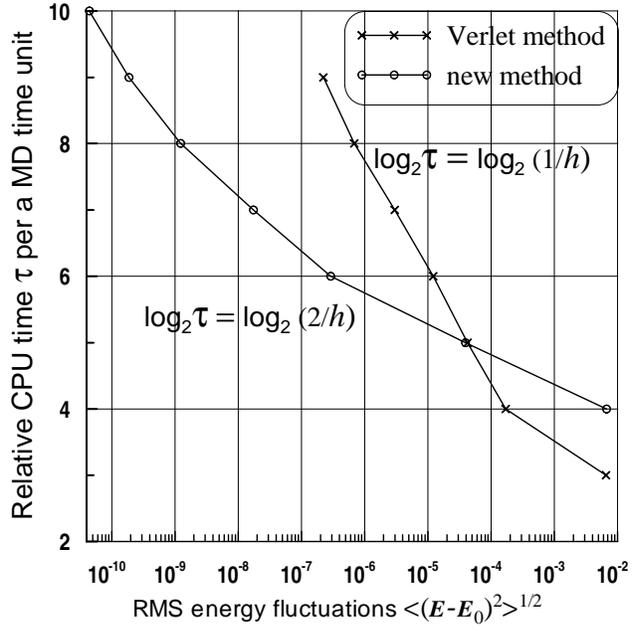}
 \caption{\label{fig:Eff-Err} Comparison of the relative CPU times $\tau$ required for simulation of 1 MD time
                              unit at given accuracy. It is assumed the $\tau=1$ at time-step $h=1$ for
                              Verlet method and the $\tau=2$ at time-step $h=1$ for a new method. }
 \end{center}
\end{figure}

We have performed a practical test of the new method to estimate its applicability for MD simulation problem. The pair
potential for our model system is described by the quasi-Lennard-Jones (qLJ) pair potential with cut-off distance
$r_{c}^2=5 r_0^2$, where $r_{0}^6= 2$. In reduced MD units with LJ parameters $\sigma=1$,$\epsilon=1$ it is given by
\begin{equation}
  \phi_{qLJ}(r) = \frac{ r_0^2}{2^8}\left(r^2- r_c^2 \right)^4 \left[r^{-10}-r^{-4} \right]      \label{eq:LJ1}
\end{equation}
Unlike to standard LJ potential function the qLJ potential Eq.(\ref{eq:LJ1}) has "good" properties at cut-off radius
$r=r_c\approx 2.51$, resulting in better energy conservation. The atom mass is assumed to be $m=48$. Number of atoms is
fixed $N=3375$ and all of them are located inside the MD simulation cube $L=17$ MD units with imposed periodical
conditions. Before testing the MD system is thermalized at $T=1.2$ and stored on disk. This thermodynamically
equilibrium state is a starting point of all testing runs.

For comparison purposes we choose Verlet algorithm in the coordinate form Eqs.(\ref{eq:Verl_d},\ref{eq:Verl_x}):
\begin{eqnarray}
   \delta x_h = \delta x_0 + a_0h^2 + O(h^4)    \qquad \qquad                 & &  \label{eq:Verl_d}     \\
          x_h = x_0 + \delta x_h         \qquad \qquad \qquad \qquad \quad    & &  \label{eq:Verl_x}     \\
          a_h = f(x_h)/m                 \qquad \qquad \qquad \qquad \quad    & &  \label{eq:Verl_a}     \\
          v_h = (\delta x_h+\delta x_0)/2h + (a_{h} + 2a_0)h/3 + O(h^4)       & &  \label{eq:Verl_v}
\end{eqnarray}
To improve the local energy conservation in the Verlet scheme the particle velocity $v_h$ is evaluated by using
accurate Eq.(\ref{eq:Verl_v}) which does not affect the trajectory evaluation (see \cite{Zhakh97}).

The new integrator is chosen in the form of Eqs.(\ref{eq:idea12a}, \ref{eq:2Ex5}, \ref{eq:2Ex6}) with step-by-step
algorithm from the Table 2. Unfortunately, the new method requires, in principle, the double run over all $i \in
\{1,..,N\}$ particles to evaluate $\mb{a}_{i}$ and $\mb{c}_{i} = \sum \mb{c}_{ij}$. According to our simulations it
takes approximately 2 (from 1.8 to 2.2 on different machines) times more computer time than it takes for the Verlet
scheme.

Figures 1,2 show the timing data of several simulation runs for different time steps. Fig.1 indicates
that root-mean-square accuracy of the new method is the superior, but the new method takes almost twice
as much of CPU time. We evaluate the relative efficiency of the method by comparing timings required for simulation
of the same MD time by the new method and Verlet method at given accuracy. The results are demonstrated in Fig.2 .
For given test the new integrator becomes more efficient if the desired accuracy is better than
$5\times 10^{-5}$. For instance, at the prescribed accuracy $10^{-6}$ the new method takes 4 times less computer
time than the Verlet algorithm.

\section*{ CONCLUSION }

We have demonstrated that the new integrator is time-reversible and very accurate. It can be efficiently applied for
highly precise MD simulations. The new method is expected to be useful for simulation of celestial mechanics problems
too.

It should be noted that Eq.(\ref{eq:2Ex6}) was obtained first in \cite{Toxvaerd}, but it was used in
a time-irreversible algorithm.

Author thankfully acknowledges Prof.Nishihara for his kind invitation on guest-professor position at ILE, Osaka
University in summer season, 2003. The work was supported by the Japan Society for Promotion of Science, ID No.L03541.


\end{document}